\def\gsim{\mathrel{\rlap{\lower4pt\hbox{\hskip1pt$\sim$}}\raise1pt\hbox{$>$}}}
\documentclass{article}
\usepackage{latexsym}
\usepackage{graphicx}
\begin{document}
\begin{center}
{\large \textit{pp}$\;$}{\large \textbf{Elastic Scattering at LHC in Near Forward Direction}}
\end{center}
\begin{center}
{M.M. ISLAM$^*$ and R.J. LUDDY$^{\dagger}$\\
\small \textit{{Department of Physics, University of Connecticut, Storrs, CT 06269, USA}}}\\
\small \textit{{$^*$islam@phys.uconn.edu \hspace{1cm}$^{\dagger}$rjluddy@phys.uconn.edu}}\\
\vspace{.1cm}
{A.V. PROKUDIN\\
\small \textit{{Dipartmento di Fisica Teorica, Universit$\hat{a}$
Degli Studi di Torino, Via Pietro Giuria 1,\\
10125 Torino, Italy and Sezione INFN di Torino, Italy\\
Institute for High Energy Physics, 142281 Protvino, Russia\\
prokudin@to.infn.it}}}
\end{center}
\begin{abstract}
We predict \textit{pp} elastic differential cross section at LHC at the c.m. energy 
$\sqrt s $ = 14 TeV and momentum transfer range $\vert t\vert $ = 0 -- 10 
GeV$^{2}$, which is planned to be measured by the TOTEM group. The field 
theory model underlying our phenomenological investigation describes the 
nucleon as a composite object with an outer cloud of quark-antiquark 
condensate, an inner core of topological baryonic charge, and a still 
smaller quark-bag of valence quarks. The model satisfactorily describes the 
asymptotic behavior of $\sigma _{tot}$(s) and $\rho $(s) as well as the 
measured $\bar {p}p$ elastic d$\sigma $/dt at $\sqrt s $ = 546 GeV, 630 GeV, 
and 1.8 TeV. The large $\vert t\vert $ elastic amplitude of the model 
incorporates the BFKL Pomeron in next to leading order approximation, the 
perturbative dimensional counting behavior, and the confinement of valence 
quarks in a small region within the nucleon.

\vspace{0.2cm}

\end{abstract}
\textit{pp} elastic differential cross section at LHC in near forward direction at c.m.
energy $\sqrt s $ = 14 TeV and momentum transfer $\vert t\vert $ = 
0 -- 10 GeV$^{2}$ is planned to be measured by the TOTEM (TOTal and 
Elastic Measurement) group [1]. Various models have been proposed to 
describe \textit{pp} elastic scattering in the diffraction region $\vert t\vert $ 
$\simeq$ 0 - 0.5 GeV$^{2}$, such as: i) single Pomeron exchange with a 
trajectory $\alpha _{P}$(t)=1.08+0.25t [2], ii) multiple Pomeron exchanges 
with single- and double- diffractive dissociation[3], iii) the incident 
proton viewed as made-up of two color dipoles in the target proton rest 
frame[4]. \textit{pp} elastic d$\sigma $/dt at LHC all the way from $\vert t\vert $ = 
0 to 10 GeV$^{2}$ has been predicted on the basis of three different models: 
a) impact-picture model [5] based on the Cheng-Wu calculation of QED tower 
diagrams [6], b) eikonalized Pomeron-Reggeon model using conventional Regge 
approach, but with multiple Pomeron-Reggeon exchanges included [7,8], c) 
effective field theory model that describes the nucleon as a chiral-bag with 
a quark-antiquark cloud [9,10]. A QCD motivated eikonalized model has also 
been proposed to predict \textit{pp} d$\sigma $/dt at $\sqrt s $ = 14 TeV for $\vert 
 t \vert $ = 0 -- 2.0 GeV$^{2}$ [11]. This wide array of models attempting 
to describe $pp$ elastic scattering at LHC reflects the view that 
quantitative understanding of this process will provide fundamental insight 
into the nonperturbative and the perturbative QCD dynamics.

The impact-picture model and the eikonalized Pomeron-Reggeon model predict 
besides the first dip-bump structure more diffraction-like secondary 
structures at large $\vert t \vert $ [5, 7, 8]. The chiral-bag model with 
$q\bar {q}$ condensate cloud, which we studied [9], predicts after the first 
dip-bump structure a smooth approximately exponential fall-off (known as 
Orear fall-off) and then a slower fall-off due to the transition from the 
nonperturbative regime to the perturbative regime. This change in the 
behavior of d$\sigma $/dt was shown only schematically in our previous work 
[9]. We have now been able to quantitatively address this question and study 
the predicted change of d$\sigma $/dt. Results of our investigation and the 
implications for the combined role of perturbative and nonperturbative QCD 
dynamics are briefly reported here.

We view \textit{pp} elastic scattering in the perturbative regime as a hard collision 
of a valence quark from one proton with a valence quark from the other
proton (Fig.1). The collision carries off the whole
\begin{figure}[ht]
  \hspace{0.6cm}\includegraphics[height=1.7in]{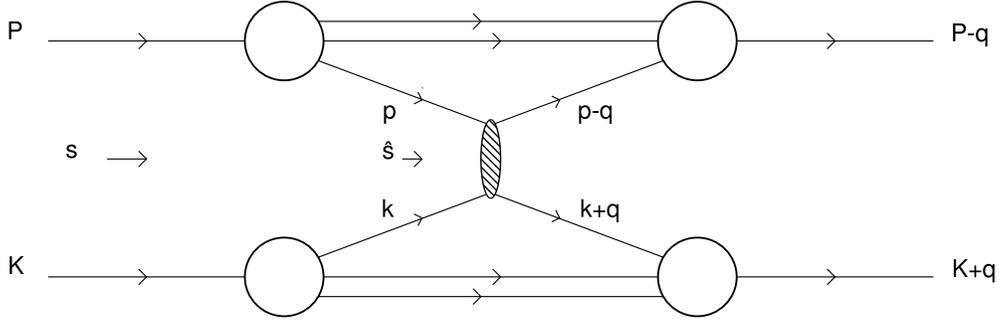}
  \caption{\small{Hard collision of valence quarks from two different protons}}
\end{figure}
momentum transfer. This dynamical picture brings new features in our 
calculations: $\;$ 1)Probability amplitude of a quark to have, say, momentum 
$\vec {p}$ when the proton has momentum $\vec {P}$ in the c.m. frame. 2) 
Quark-quark elastic amplitude at high energy and large momentum transfer, 
which is in the domain of perturbative QCD. The latter has been the focus of 
extensive studies following the original work of Balitsky, Fadin, Kuraev, 
and Lipatov (BFKL) [12]. The present status is that the \textit{qq} elastic scattering 
occurs via Reggeized gluon ladders with rungs of gluons which represent 
gluon emissions in inelastic processes (BFKL Pomeron). It is a crossing-even 
amplitude which is a cut in the angular momentum plane with a fixed branch 
point at $\alpha _{BFKL}$ = 1+$\omega $. The value of $\omega $ in the 
next-to-leading order (NLO) lies in the range 0.13-0.18 as argued by Brodsky et al.[13]. We refer to 
the BFKL Pomeron with next to leading order corrections included as the QCD 
``hard Pomeron''. In our investigation, we approximate this hard Pomeron by 
a fixed pole and take the \textit{qq} scattering in Fig. 1 as
\begin{equation}
\label{eq:1a}
\hat{T}(\hat{s},t)=i\gamma _{qq}\hat{s}\,(\hat{s}\;e^{-i\frac{\pi}{2}})^\omega \;\frac{1}{\vert t\vert +r_0^{-2}} ,
\end{equation}
where $\hat {s}=(p+k)^2,\;\;t=-\vec {q}^{\;2}\,.$ The phase in Eq.(\ref{eq:1a}) follows
from the requirement that $\hat {T}(\hat {s},t)$ is a crossing even 
amplitude. Eq.(\ref{eq:1a}) represents the hard Pomeron amplitude in our 
calculations. If we want to describe just asymptotic \textit{qq} scattering, we have to 
take into account unitarity corrections due to infinite exchanges of this 
Pomeron. This can be done by taking $\hat {T}(\hat {s},t)$ as the Born 
amplitude in an eikonal formulation [14], which leads to a black-disk 
description and requires $\gamma _{qq} >$ 0. The radius of the black disk 
turns out to be $R(\hat {s})=r_0 \omega \ln \hat {s}.$ Hence, the parameter 
$r_0 $ in Eq.(\ref{eq:1a}) has the physical significance of a length scale that 
defines the black-disk radius of asymptotic quark-quark scattering.

We next examine how to obtain the \textit{pp} elastic scattering amplitude from the 
process shown in Fig. 1. Let $s$ be the square of the c.m. energy of the two 
colliding protons: $s=(P+K){ }^2. \; \hat{s},\; $of course, is the square of 
the c.m. energy of the two colliding quarks. From Fig. 1, we see that 
initially we have a quark of momentum $\vec {p}:\;\left| {\vec {p}} 
\right\rangle $ with a probability amplitude $\varphi (\vec {p})$ in the 
c.m. frame in which the proton is moving with momentum $\vec {P}.$  Similarly, we have a second quark with momentum $\vec {k}:\;\left| {\vec 
{k}} \right\rangle $ with a probability amplitude $\varphi (\vec {k})$ in 
the c.m. frame in which the other proton is moving with momentum $\vec 
{K}=-\vec {P}.$ Thus, the initial state of the two colliding quarks is 
\begin{equation}
\label{eq:2a}
\left| i \right\rangle =\varphi (\vec {p})\left| {\vec {p}} \right\rangle 
\varphi (\vec {k})\left| {\vec {k}} \right\rangle \;.
\end{equation}
After the collision, we have a quark with momentum $\vec {p}-\vec 
{q}:\;\left| {\vec {p}-\vec {q}} \right\rangle $ with a probability 
amplitude $\varphi (\vec {p}-\vec {q})$, and a quark with momentum $\vec 
{k}+\vec {q}:\;\left| {\vec {k}+\vec {q}} \right\rangle $ with a probability 
amplitude $\varphi (\vec {k}+\vec {q}).$ So, the final state is 
\begin{equation}
\label{eq:3a}
\left| f \right\rangle =\varphi (\vec {p}-\vec {q})\left| {\vec {p}-\vec 
{q}} \right\rangle \varphi (\vec {k}+\vec {q})\left| {\vec {k}+\vec {q}} 
\right\rangle \;.
\end{equation}
The \textit{pp} elastic scattering amplitude due to quark-quark scattering $T_{qq} 
(s,-\vec {q}^{\;2})$ from Fig. 1 is then 
\begin{equation}
\label{eq:4a}
T_{qq} (s,-\vec{q}^{\;2})=\sum\limits_{\vec {p}} \sum\limits_{\vec {k}} 
{\varphi ^\ast (\vec {p}-\vec {q})\varphi ^\ast (\vec {k}+\vec 
{q})\left\langle {\vec {k}+\vec {q}} \right|\left\langle {\vec {p}-\vec {q}} 
\right|\hat {T}_{op} \left| {\vec {p}} \right\rangle \left| {\vec {k}} 
\right\rangle } \varphi (\vec {p})\varphi (\vec {k})\;,
\end{equation}
where $\left\langle {\vec {k}+\vec {q}} \right|\left\langle {\vec {p}-\vec 
{q}} \right|\;\hat {T}_{op} \;\left| {\vec {p}} \right\rangle \left| {\vec 
{k}} \right\rangle $ is the \textit{qq} elastic scattering amplitude. Since this 
amplitude only depends on the invariants $\hat {s}=(p+q)^2$ and $\hat 
{t}=-\vec{q}^{\;2}\;,$ we can write
\begin{equation}
\label{eq:5a}
\left\langle {\vec {k}+\vec {q}} \right|\left\langle {\vec {p}-\vec {q}} 
\right|\;\hat {T}_{op} \;\left| {\vec {p}} \right\rangle \left| {\vec {k}} 
\right\rangle =\hat {T}(\hat {s},-\vec {q}^{\;2}).
\end{equation}
Eq.(\ref{eq:4a}) then takes the form 

\begin{equation}
\label{eq:6a}
T_{qq} (s,-\vec {q}^{\;2})=\sum\limits_{\vec {p}} \sum\limits_{\vec {k}} 
{\varphi ^\ast (\vec {p}-\vec {q})\varphi (\vec {p})\;\hat {T}(\hat 
{s},-\vec {q}^{\;2})\;\varphi ^\ast (\vec {k}+\vec {q})} \varphi (\vec {k})\;.
\end{equation}
This equation makes it evident that $\varphi ^\ast (\vec {p}-\vec 
{q})\varphi (\vec {p})$ and $\varphi ^\ast (\vec {k}+\vec {q})\varphi (\vec 
{k})$ are the nonperturbative ``impact factors'' which modify the 
perturbative \textit{qq} amplitude $\hat {T}(\hat {s},-\vec {q}^{\;2}).$ The 
right-hand-side (RHS) of Eq.(\ref{eq:6a}) needs to be multiplied by a factor of nine 
to take into account that there are three quarks in each proton[15]. We 
absorb this factor in the constant $\gamma _{qq} $.

To see the physical meaning of Eq.(\ref{eq:6a}), let us assume that we can 
approximate \textit{qq} scattering in Fig.1
by taking some average value of $\hat {s}:\hat {s}_{av} .$ Of course, $\hat 
{s}_{av} $ is going to be proportional to $s$. Eq.(\ref{eq:6a}) then takes the form 
\begin{equation}
\label{eq:7a}
T_{qq} (s,-\vec {q}^{\;2})\simeq \sum\limits_{\vec {p}} {\varphi ^\ast (\vec 
{p}-\vec {q})\varphi (\vec {p})\;\hat {T}(\hat {s}_{av} ,-\vec 
{q}^{\;2})\sum\limits_{\vec {k}} {\varphi ^\ast (\vec {k}+\vec {q})\varphi (\vec 
{k})} } \;,
\end{equation}
which shows that the impact factors separate out. Each momentum sum in
Eq.(\ref{eq:7a})
can now be carried out and yields the form factor associated with the quark 
probability density in the c.m. frame. This probability density is Lorentz 
contracted, which means if $\rho _0 (\vec {{r}'})$ is the quark probability 
density at $\vec {{r}'}$ in the proton
rest frame and $\rho (\vec {r})$ is the probability density at $\vec {r}$ in 
the c.m. frame, then
\begin{equation}
\label{eq:8a}
\rho (\vec {b}+\vec {e}_3 z)=\gamma \,\rho _0 (\vec {b}+\vec {e}_3 \gamma 
{\kern 1pt}z),
\end{equation}
where $\gamma $ is the Lorentz contraction factor: $\gamma =E/M={\sqrt s } 
\mathord{\left/ {\vphantom {{\sqrt s } {(2M)}}} \right. 
\kern-\nulldelimiterspace} {(2M)}$, $\vec {r}=\vec {b}+\vec {e}_3 z$, and 
$\vec {e}_3 $ is the unit vector in the direction of $\vec {P}$, i.e., the
z-axis. If $F(\vec {q})$ is the form factor associated with $\rho _0 (\vec {r})$: 
\begin{equation}
\label{eq:9a}
F(\vec {q})=\int {d^3r\;e^{i\vec {q}\cdot \vec {r}}\rho _0 (\vec {r})} ,
\end{equation}
and $\rho _0 (\vec {r})$ is spherically symmetric, then
\[
\sum\limits_{\vec {p}} {\varphi ^\ast (\vec {p}-\vec {q})\varphi (\vec 
{p})\;=} \int {d^3r\;e^{-i\vec {q}\cdot \vec {r}}\rho (\vec {r})} 
\]
\begin{equation}
\label{eq:10a}
\quad \quad \quad \quad \quad \quad \quad =F(\vec {q}_\bot +\vec {e}_3 
\frac{q_3 }{\gamma }).
\end{equation}
In deriving Eq.(\ref{eq:10a}), we have used $\rho (\vec {r})=\psi ^\ast (\vec {r})\psi 
(\vec {r})$, where the quark wave function $\psi (\vec {r})$is related to 
its momentum wave function $\varphi (\vec {p})$ via the plane wave 
expansion: 
\begin{equation}
\label{eq:11a}
\psi (\vec {r})=\sum\limits_{\vec {p}} {\frac{e^{i\vec {p}\cdot \vec 
{r}}}{\sqrt V }\varphi (\vec {p})} 
\end{equation}
Eq.(\ref{eq:7a}) now takes the form 
\begin{equation}
\label{eq:12a}
T_{qq} (s,-\vec {q}^{\;2})\simeq F(\vec {q}_\bot )\;\hat {T}(\hat {s}_{av} 
,-\vec {q}^{\;2})\;F(\vec {q}_\bot ),
\quad
(\frac{q_3 }{\gamma }=\frac{2Mq_3 }{\sqrt s }\to 0).
\end{equation}
The structure of Eq. (\ref{eq:12a}) is easy to understand. It is the usual 
quantum-mechanical scattering amplitude of two composite objects described 
by the form factors and interacting via a basic process whose amplitude is 
$\;\hat {T}(\hat {s}_{av} ,-\vec {q}^{\;2})$. We take the form factor $F(\vec {q})$ describing the quark probability density or number density in 
the nucleon rest frame to be a dipole:
\begin{equation}
\label{eq:13a}
F(\vec {q})=\left( {\,1+\textstyle{{\vec {q}^{\;2}} \over {m_0 ^2}}} 
\right)^{\,-2}\quad ,
\end{equation}
so that it satisfies the dimensional counting behavior $t^{ -2}$ for the form 
factor of a proton made up of three quarks [16,17,18].

Next we go back to Eq.(\ref{eq:6a}) and no longer make the approximation of replacing 
$\hat {s}$ by an average value. Inserting Eq.(\ref{eq:1a})
in Eq.(\ref{eq:6a}), we obtain
\begin{equation}
\label{eq:14a}
T_{qq} (s,-\vec{q}^{\;2})=\sum\limits_{\vec {p}} \sum\limits_{\vec {k}} 
{\varphi ^\ast (\vec {p}-\vec {q})\varphi (\vec {p})\;
i\gamma _{qq}\hat{s}\,(\hat{s}\;e^{-i\frac{\pi}{2}})^\omega \;
\frac{1}{\vec {q}^2+r_0^{-2} }\;} \varphi ^\ast (\vec {k}+\vec {q})\varphi 
(\vec {k})\;.
\end{equation}
Introducing light-cone variables 
$P_+ =P_0 +P_3 \;,P_- =P_0 -P_3 \;,p_+ =p_0 +p_3 \;,p_- =p_0 -p_3 \;,$etc. and 
writing $p_+ =x\,P_+ \;,\;k_- =x'\,K_- \;,$we find $\hat {s}\;\simeq  
x\,x'\,s\;,$ when $P_+ \,,\,K_- \to \infty \;.$ Eq.(\ref{eq:14a}) then takes the 
separable form
\begin{equation}
\label{eq:15a}
T_{qq} (s,-\vec {q}^{\;2})=\left( {\sum\limits_{\vec {p}} {\varphi ^\ast (\vec 
{p}-\vec {q})\varphi (\vec {p})\,x^{1+\omega }} } \right)\;
i\gamma _{qq}s\,(s\; e^{-i\frac{\pi}{2}})^\omega
\;\frac{1}{\vec {q}^{\;2}+r_0^{-2} }\;\left( {\sum\limits_{\vec {k}} {\varphi
^\ast (\vec {k}+\vec {q})\varphi (\vec{k})\,x'\,^{1+\omega }} } \right).
\end{equation}

In a frame where $P_+ \,\to \infty \;,$
\begin{equation}
\label{eq:16a}
\sum\limits_{\vec {p}} {\varphi ^\ast (\vec {p}-\vec {q})\varphi (\vec 
{p})\,x^{1+\omega }} =\frac{M\,m_0^5 }{8\pi }\int\limits_0^1 {dx} 
\frac{x^{1+\omega }}{\left( {\frac{m_0^2 }{4}+M^2x^2} \right)}\;I(q_\bot 
,\alpha (x))\;,
\end{equation}
where
\begin{equation}
\label{eq:17a}
I(q_\bot ,\alpha (x))\;\equiv \int\limits_0^\infty {b\,db\,J_0 (bq_\bot 
)\left\{ {bK_1 [b\alpha ]} \right\}^2} \;.
\end{equation}
Here $M$ is the nucleon mass, $m_0$ is the mass parameter that occurs in the 
form factor Eq.(\ref{eq:13a}), $\alpha =\left( {\frac{m_0^2 }{4}+M^2x^2} 
\right)^{\textstyle{1 \over 2}}\;$, and $\vec {q}\simeq \vec {q}_\bot 
\,.$ In deriving Eq. (\ref{eq:16a}), we use momentum wave function $\varphi
(\vec {p})$ obtained from the Lorentz contracted probability density.
It can be related to the rest frame wave function $\varphi _0 ({\vec {p}}')$ 
in the following way:
\begin{equation}
\label{eq:18a}
\varphi (\vec {p}_\bot +\vec {e}_3 p_3 )=\varphi _0 (\vec {p}_\bot +\vec 
{e}_3 \frac{p_3 }{\gamma })\;,
\end{equation}
and yields the result 
\begin{equation}
\label{eq:19a}
\varphi (\vec {p}_\bot +\vec {e}_3 p_3 )=
\left( {\frac{2\pi \;m_0^5 }{V_0 }}
\right)^{\frac{1}{2}}
\left( \frac{m_0^2 }{4}+p_\bot ^2 +\frac{p_3^2 }{\gamma ^2} \right)^{-2}.
\end{equation}
($V_0$ is the quantization volume in the rest frame.)
The integral $I(q_\bot ,\alpha (x))$ can be evaluated analytically, and we
obtain
\begin{equation}
\label{eq:20a}
I(q_\bot ,\alpha (x))=\frac{1}{8\alpha ^4}\left\{ {\frac{2}{a^3a'}\ln 
(a'+a)+\frac{1}{aa'^3}\ln (a'+a)-\frac{1}{a^2a'^2}-\frac{3a'}{a^5}\ln 
(a'+a)+\left. {\frac{3}{a^4}} \right\} } \right.,
\end{equation}
where $a'^2=\frac{q_\bot ^2 }{4\alpha ^2}\;,\;a^2=a'^2+1\;.$
Let us denote by $\mathcal{F}(q_\bot )$ the RHS of Eq.(\ref{eq:16a}).
Eq.(\ref{eq:15a}) can then be expressed in the form
\begin{equation}
\label{eq:21a}
T_{qq} (s,-\vec{q}^{\;2})=\mathcal{F}(q_\bot )\;
i\;\gamma _{qq}s\,(s\; e^{-i\frac{\pi}{2}})^\omega \;
\frac{1}{\vert t\vert +r_0^{-2} }\mathcal{F}(q_\bot )\;.
\end{equation}
For $a'^2=\frac{q_\bot ^2 }{4\alpha ^2}>>1$ and $a^2\simeq a'{ }^2$,
Eq.(\ref{eq:20a}) yields
\begin{equation}
\label{eq:22a}
I(q_\bot ,\alpha (x))\;\simeq \;\frac{4}{q_\bot ^4 }\;\;\simeq  
\frac{4}{\vert t\vert ^2}\;,\quad \quad (\,\vert t\vert =\vec {q}^{\;2} \simeq 
 q_\bot ^2 ).
\end{equation}
Substituting this on the RHS of Eq.(\ref{eq:16a}), we find
\begin{equation}
\label{eq:23a}
\mathcal{F}(q_\bot )\sim \frac{1}{\vert t\vert ^2}\;.
\end{equation}
Eq.(\ref{eq:21a}) then leads to an amplitude
\begin{equation}
\label{eq:24a}
T_{qq} (s,-\vec {q}^{\;2})\sim \;
\frac{i\;\gamma_{qq}s\,(s\;e^{-i\frac{\pi}{2}})^\omega}
{\vert t\vert ^5}\;.
\end{equation}
This results in differential cross section behavior for fixed $s$ and large 
$\vert t \vert $: 
\begin{equation}
\label{eq:25a}
\frac{d\sigma }{dt}\sim \frac{1}{\vert t\vert ^{10}}\;,
\quad
(s>>\vert t\vert >>m_0^2 +4M^2).
\end{equation}
Eq.(\ref{eq:25a}) shows that we obtain the behavior 
predicted by the perturbative QCD dimensional counting rules[16,17,18] for 
large $\vert t \vert $.

In our \textit{pp} elastic scattering model, we now have two hard-collision amplitudes: 
one due to $\omega $ exchange, the other due to the hard Pomeron exchange. 
Both collisions are accompanied by cloud-cloud diffraction scattering that 
reduces these amplitudes by an absorption factor $\exp (i\hat {\chi }(s,0))$ 
[19]. So the sum of the two hard amplitudes becomes
\begin{equation}
\label{eq:26a}
T_1 (s,t)=\,\;e^{i\hat {\chi }(s,0)}\left[ {\pm \tilde {\gamma 
}\;s\frac{F^2(t)}{m^2-t}+
i\;\gamma _{qq}s\,(s e^{-i\frac{\pi}{2}})^\omega 
\;\frac{\mathcal{F}^2(q_\bot )}{\vert t\vert +r_0^{-2} }} \right]
\;, (+ for \; \bar{p}p, - for\; pp).
\end{equation}
Using the same parameterization as before [9],
\begin{equation}
\label{eq:27a}
\tilde {\gamma }\;e^{i\hat {\chi }(s,0)}=\hat {\gamma }_0 \;+\;\frac{\hat 
{\gamma }_1 }{(se^{-i\frac{\pi }{2}})^{\hat {\sigma }}}\;,
\end{equation}
we find 
\begin{equation}
\label{eq:28a}
T_1 (s,t)=\,\;\left[ {\hat {\gamma }_0 \;+\;\frac{\hat {\gamma }_1 
}{(s\;e^{-i\frac{\pi }{2}})^{\hat {\sigma }}}} \right]\;\left[ {\pm 
\;s\frac{F^2(t)}{m^2-t}+
i\;\tilde{\gamma}_{qq}s\,(s\;e^{-i\frac{\pi}{2}})^\omega 
\;\frac{\mathcal{F}^2(q_\bot )}{\vert t\vert +r_0^{-2} }} \right],
\end{equation}
where $\tilde {\gamma }_{qq} =\;\gamma _{qq} /\tilde {\gamma }$. The \textit{qq} hard 
scattering term brings four new parameters: i) $\tilde {\gamma }_{qq} $which 
measures the relative strength of this term compared to the $\omega $ 
exchange term; ii) $\alpha _{BFKL}$ = 1+$\omega $ which controls the high 
energy behavior; iii) $r_0 $ which provides the length scale for the 
black-disk radius of \textit{qq} asymptotic scattering; iv) $m_0 $ which determines the 
quark wave function $\psi _0 (\vec {r})=\sqrt {\rho _0 (\vec {r})} $ and the 
size of the quark bag. Because of the different physical aspects associated 
with them, these four parameters form a minimal set.

We determine the parameters of the model by requiring that the model should 
describe satisfactorily the asymptotic behavior of $\sigma _{tot} (s)\;$and 
$\rho (s)$ as well as the measured $\bar {p}p$ elastic d$\sigma $/dt at 
$\sqrt s $=546 GeV
[20], 630 GeV [21], and 1.8 TeV [22, 23]. The results of this investigation 
are shown in Figs. 2 - 4 together with the experimental data. We obtain 
quite satisfactory descriptions. The dotted curves in Figs. 2 and 3 
represent the error bands given by Cudell et. al. (COMPETE Collaboration) to 
their best fit [24]. We notice that our $\sigma _{tot} (s)\;$curve lies within
their error band closer to the lower curve, but our $\rho _{pp} (s)$ curve (dashed
curve in Fig.3) deviates from the band. As noted by Cudell et. al., such a deviation is not surprising -- since 
a hard Pomeron occurs in our calculations and not in theirs. In fact, this 
hard Pomeron in conjunction with a crossing-odd absorptive correction [19] 
in our model leads to a crossing-odd amplitude (an odderon) and produces a visible 
difference between $\rho _{\bar {p}p} (s)$ and $\rho _{pp} (s)$ at large 
$\sqrt s $. The parameters describing the soft 
(small $\vert t \vert )$ diffraction amplitude and the hard (large $\vert 
 t \vert )\; \omega $-exchange amplitude have been discussed before [9]. 
Their values are:
$R_0 =2.77, R_1 =0.0491, a_0 =0.245, a_1 =0.126, \eta _0 =0.0844, c_0 =0.00,
\sigma =2.70, \lambda_0 =0.727, d_0 =13.0, \alpha =0.246,
\hat {\gamma }_0 =1.53, \hat {\gamma }_1 =0.00, \hat {\sigma }=1.46$
(the unit of energy is 1 GeV). The parameters $\beta$ and $m$ are kept fixed
as previously: $\beta$=3.075, $m$=0.801. There are now seventeen adjustable parameters.
The four new parameters describing the
hard (large $\vert t \vert )$ \textit{qq} amplitude have the
values $\tilde {\gamma }_{qq} =0.03$, $\omega =0.15$, $r_0 =2.00$,
$m_0^2 =12.0.$ (This value of $m_0^2 $ leads to a valence quark-bag of r.m.s. 
radius 0.2 F, while that of the baryonic charge core is 0.44 F.) These four
parameters, however, cannot be determined reliably, because no large
$\vert t \vert$ elastic data are available in the TeV energy region.

Our prediction for \textit{pp} elastic differential cross section at LHC at $\sqrt s $= 
14 TeV for the whole momentum transfer range $\vert t \vert $ = 0 -- 10 
GeV$^{2}$ is now given in Fig. 5 (solid curve). We obtain for $\sigma _{tot} 
$ and $\rho _{pp} $ the values 110 mb and 0.120 respectively. Also given in 
Fig. 5 are separate d$\sigma $/dt due to diffraction (dotted curve), due to 
hard $\omega $-exchange (dot-dashed curve), and due to hard \textit{qq} scattering 
(dashed curve). As expected in our model, we find that in the small $\vert 
 t \vert $ region ($\vert t \vert \simeq $0 -- 0.5 GeV$^{2})$ diffraction 
dominates, in the intermediate $\vert t \vert $ region ($\vert t \vert \simeq$ 
1.0 -- 4.0 GeV$^{2}) \quad \omega $-exchange dominates, and in the
large $\vert t \vert $ region
($\vert t \vert $
$\gsim$
 6.0 GeV$^{2})$ \textit{qq} scattering dominates.
The three $\vert t \vert $ regions correspond to cloud-cloud interaction, 
core-core scattering due to $\omega $-exchange, and valence \textit{qq} scattering via 
QCD hard Pomeron. Therefore, they reflect the composite structure of the 
nucleon with an outer cloud, an inner core of topological baryonic charge, 
and a still smaller quark-bag of valence quarks.

We note that \textit{pp} elastic differential cross section in the energy range $\sqrt 
s $= 27 - 62 GeV and $\vert t \vert \ge $ 3.5 Gev$^{2}$ was observed to 
be approximately energy independent and falling off as $t ^{-8}$. This was 
interpreted as due to the independent exchanges of three perturbative gluons 
[25, 26]. Later it was pointed out that the three gluons would Reggeize, so 
that color-octet exchanges would be suppressed. Instead, three color-singlet 
exchanges would take their place [27]. Eventually, as $\vert t \vert $ 
increases, a single color-singlet exchange would dominate and lead to a $t ^{-10}$ 
fall-off as predicted by the perturbative QCD dimensional counting rules 
[16,17,18]. In our model, the dimensional counting behavior $t ^{-10}$ of 
d$\sigma $/dt originates from the hard \textit{qq} amplitude in Eq. (28). This 
amplitude leads to a distinct change in the slope of the differential cross 
section from the intermediate $\vert t \vert $ region to the large $\vert 
 t \vert $ region as seen in Fig. 5. For example, for 1.0 $\le$ $\vert t \vert 
$ $\le$ 3.0 GeV$^{2}$, d$\sigma $/dt drops by more than two orders of magnitude, 
while for 7.0 $\le$ $\vert t \vert \le$ 9.0 GeV$^{2}$, d$\sigma $/dt drops by a 
factor of 4.2, i.e. less than an order of magnitude. Similar decrease in 
d$\sigma $/dt slope was observed at ISR by De Kerret et.al. for $\vert t \vert $
$\gsim$ 6.5 GeV$^{2}$ at a much lower energy: $\sqrt s $= 53 GeV [28].
Lepage and Brodsky[18], however, pointed out that at such low energies it would
be hard to distinguish between amplitudes that lead to $t ^{-8}$ and $t ^{-10}$
asymptotic behavior.

We conclude that, if precise measurement by the TOTEM group corroborates our
predicted slow fall-off of \textit{pp} elastic d$\sigma $/dt in the large $\vert 
 t \vert $ region, then that will provide evidence for the hard \textit{qq} amplitude 
occurring in Eq. (28). This, in turn, will imply: i) presence of the QCD 
hard Pomeron, ii) perturbative QCD dimensional counting behavior at 
asymptotic $\vert t \vert $ ($>>$ 10 GeV$^{2})$, and iii) the confinement 
of valence quarks in a small region within the proton.
\vspace{.2cm}

Much of this work was done when one of us (MMI) was at the Yang Institute
for Theoretical Physics at SUNY Stony Brook on sabbatical leave. He 
wishes to thank George Sterman, Director of the Institute, and other 
colleagues there for their hospitality. He also wishes to thank Michael 
Rijssenbeek and George Sterman for stimulating discussions.

\begin{center}{\textbf{References}}
\end{center}

1. TOTEM: Technical Design Report, Jan. 2004 (CERN-LHCC-2004-002).

2. A. Donnachie, H.G. Dosch, P.V. Landshoff, and O. Nachtmann, Pomeron 
Physics and QCD, Cambridge University Press (2002).

3. V. A. Khoze, A. D. Martin, and M.G. Ryskin, Eur. Phys.J C18 (2000) 167.

4. J. Bartels, E. Gotsman, E. Levin, M. Lublinsky, and U. Maor, Phys Lett.B 
556 (2003) 114.

5. C. Bourrely, J. Soffer, and T.T. Wu, Eur. Phys.J C28 (2003) 97.

6. H.Cheng and T.T.Wu, Expanding Protons: Scattering at High Energies, MIT 
Press, Cambridge, MA (1987).

7. P. Desgrolard, M. Giffon, E. Matynov, and E. Predazzi, Eur. Phys.J C16 
(2000) 499.

8. V. A. Petrov and A.V. Prokudin, Eur. Phys.J C23 (2002) 135.

9. M. M. Islam, R. J. Luddy, and A.V. Prokudin, Mod. Phys. Lett.A 18 (2003) 
743.

10. M. M. Islam, R. J. Luddy, and A.V. Prokudin, in AIP Conference 
Proceedings Vol.698, 2003 (Intersections of Particle and Nuclear Physics, 
edited by Z. Parsa) p.142. [hep-ph/0307355]

11. M.M. Block, E.M. Gregores, F. Halzen and G. Pancheri, Phys.Rev. D60, 
(1999) 054024. 

12. J. R. Forshaw and D. A. Ross, Quantum Chromodynamics and the Pomeron, 
Cambridge University Press (1997).

13. S. J. Brodsky, V. S. Fadin, V.T. Kim, L. N. Lipatov, and G. B. 
Pivovarov, JETP Lett.70 (1999) 155.

14. A. B. Kaidalov, Regge Poles in QCD in Handbook of QCD, Vol.1, edited by 
M. Shifman and B. Ioffe (World Scientific, 2001) p. 603.

15. The quarks in our field theory model (Ref. 10) are massless effective 
color-singlet quarks and not the QCD current quarks.

16. V. A. Matveev, R. M. Muradian, and A. N. Tavkhelidze, Lett.Nuovo Cimento 
7 (1973) 719.

17. S. J. Brodsky and G. R. Farrar, Phys.Rev.Lett.31 (1973) 1153; Phys.Rev.D 
11 (1975) 1309.

18. G.P. Lepage and S.J. Brodsky, Phys.Rev.D 22 (1980) 2157.

19. M.M. Islam, B. Innocente, T. Fearnley, and G. Sanguinetti, Europhys. 
Lett. 4 (1987) 189.

20. M. Bozzo et.al., UA4 Collaboration, Phys.Lett. B147 (1984) 385; B155, 
(1985) 197.

21. D. Bernard et.al. Phys.Lett. B171 (1986) 142.

22. N. Amos et.al., Phys.Lett. B247 (1990) 127.

23. F. Abe et.al., Phys.Rev. D50 (1994) 5518.

24. J.R. Cudell et.al., Phys.Rev.Lett. 89 (2002) 201801.

25. A. Donnachie and P.V. Landshoff, Z. Phys. C 2(1979)55.

26. A. Donnachie and P.V. Landshoff, Phys. Lett. B 387(1996) 637.

27. M. G. Sotiropoulos and G. Sterman, Nucl.Phys. B 425 (1994) 489.

28. H. De Kerret et al., Phys. Lett. B 68(1977) 374.
\twocolumn{
\begin{figure}[t]
  \includegraphics[height=2.7in]{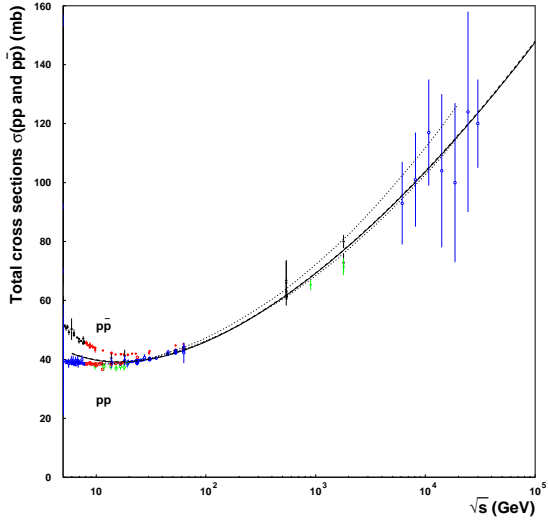}
  \caption{Solid curve represents our calculated total cross
  section as a function of $\sqrt{s}$. Dotted curves represent
  the error band given by Cudell et al. [24].}
\end{figure}
\setcounter{figure}{3}
\begin{figure}[t]
  \includegraphics[height=2.7in]{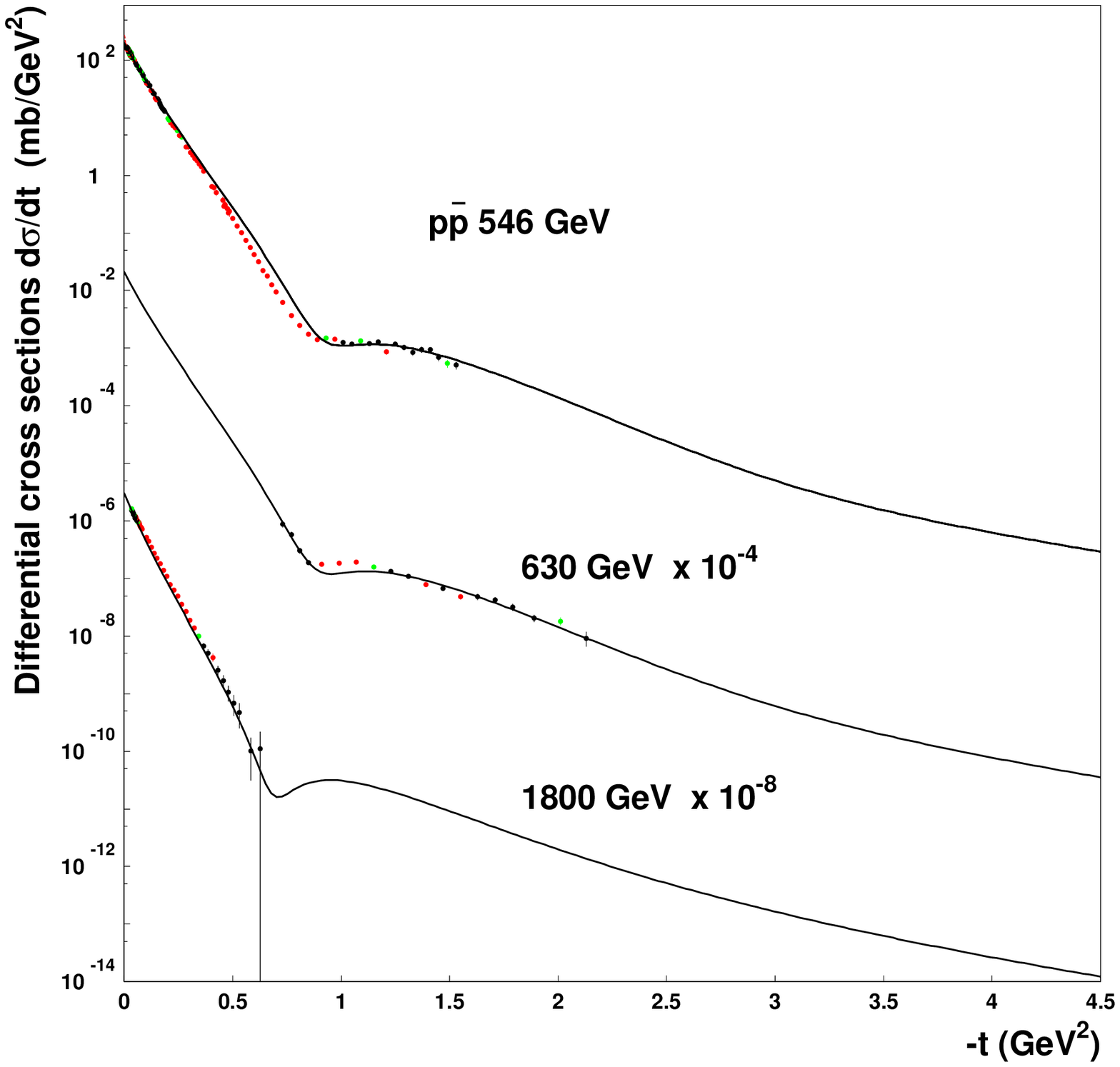}
  \caption{Solid curves show our calculated d$\sigma $/dt at
  $\sqrt{s}$ = 546, 630 and 1800 GeV. Experimental data are from
  references [20], [21] and [22, 23].\vspace{1.25cm}
  }
\end{figure}
\setcounter{figure}{2}
\begin{figure}[t]
  \includegraphics[height=2.7in]{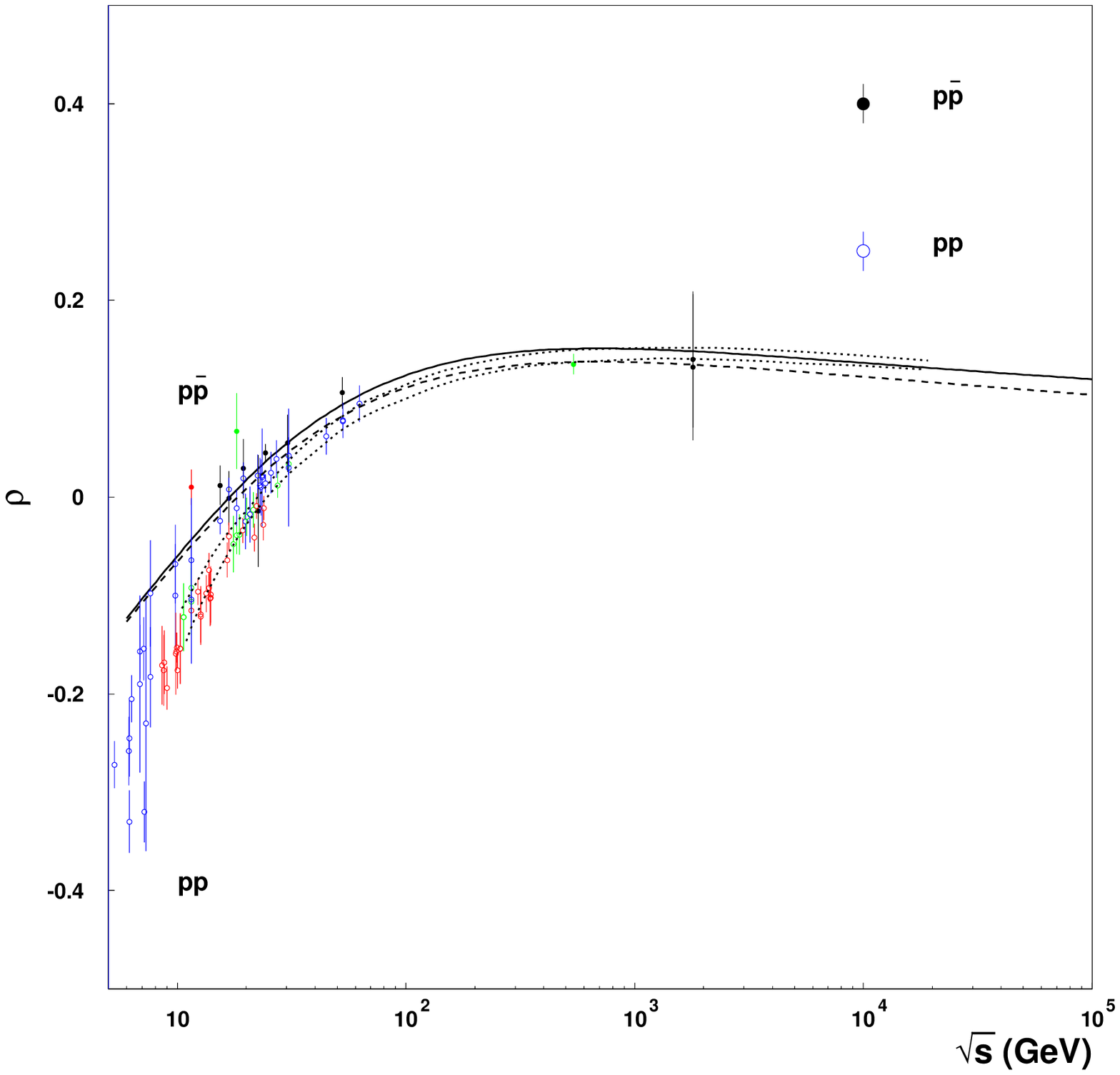}
  \caption{Solid and dashed curves represent our calculated $\rho_{\bar{p}p}$ and $\rho_{pp}$ as functions of $\sqrt{s}$.
  Dotted curves represent the error band given by Cudell et al. [24].}
\end{figure}
\setcounter{figure}{4}
\begin{figure}[t]
  \includegraphics[height=2.7in]{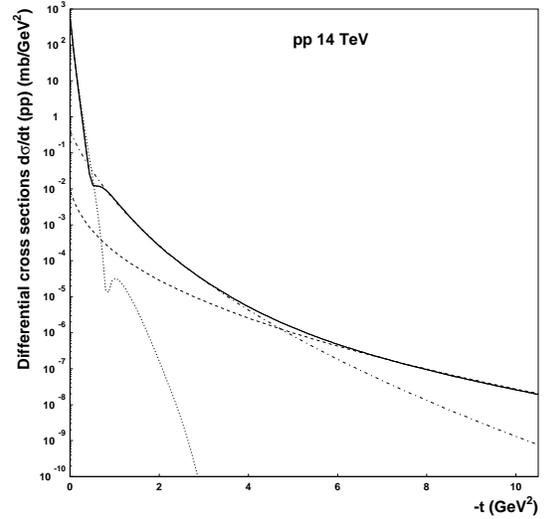}
  \caption{Solid curve shows our predicted d$\sigma $/dt for pp
  elastic scattering at $\sqrt{s}$ =14 TeV at LHC. Dotted curve
  represents d$\sigma $/dt due to diffraction only. Similarly,
  dot-dashed curve and dashed curve represent d$\sigma $/dt due to
  hard $\omega$-exchange and hard $qq$ scattering only.}
\end{figure}
}
\onecolumn
\end{document}